\documentclass[twocolumn,floatfix, showpacs]{revtex4}

\usepackage[final]{epsfig}
\usepackage{amsmath}
\usepackage{parskip}
 
\begin{document}
 
\title{Energy deposition in hard dihadron triggered events in heavy-ion collisions}
 
\author{Thorsten Renk}
\email{trenk@phys.jyu.fi}
\affiliation{Department of Physics, P.O. Box 35 FI-40014 University of Jyv\"askyl\"a, Finland}
\affiliation{Helsinki Institute of Physics, P.O. Box 64 FI-00014, University of Helsinki, Finland}
 
\preprint{HIP-2006-46/TH}

\begin{abstract}
The experimental observation of hadrons correlated back-to-back with a (semi-)hard trigger in heavy ion collisions has revealed a splitting of the away side correlation structure in a low to intermediate transverse momentum ($P_T$) regime. This is consistent with the assumption that energy deposited by the away side parton into the bulk medium produced in the collision excites a sonic shockwave (a Mach cone) which leads to away side correlation strength at large angles. A prediction following from assuming such a hydrodynamical origin of the correlation structure is that there is a sizeable elongation of the shockwave in rapidity due to the longitudinal expansion of the bulk medium. Using a single hadron trigger, this cannot be observed due to the unconstrained rapidity of the away side parton. Using a dihadron trigger, the rapidity of the away side parton can be substantially constrained and the longitudinal structure of the away side correlation becomes accessible. However, in such events several effects occur which change the correlation structure substantially: There is not only a sizeable contribution due to the fragmentation of the emerging away side parton, but also a systematic bias towards small energy deposition into the medium and hence a weak shockwave. In this paper, both effects are addressed.

\end{abstract}
 
\maketitle

\section{Introduction}

The experimental observation of hadrons correlated back-to-back with a hard or semi-hard trigger hadron in Au-Au collisions at 200 AGeV has revealed a splitting of the away side correlation peak in a semi-hard momentum regime between 1 and 2.5 GeV \cite{PHENIX_Mach,PHENIX_Mach2,STAR_Mach0,STAR_Mach} wich is absent in p-p collisions where two back-to-back peaks appear. This means that the main strength of the away side correlation in Au-Au collisions in this momentum region is not found in the direction of the away side parton  but at a large angle with respect to it. This angle is found to remain constant if the trigger momentum is changed and also for a variety of associate hadron momenta in the semi-hard regime. This observation can be contrasted with back-to-back correlations at hard trigger and associate hadron momenta well above 4 GeV \cite{STAR_Dijets} which show a reappearance of back-to-back correlations as seen in p-p collisions, albeit suppressed. 

This pattern has given rise to the idea that while energy loss of a back-to-back parton pair is responsible for the suppression observed at high $P_T$, the measurements at intermediate associate hadron $P_T$ show how this energy is redistributed into the medium and may in fact show the recoil of the medium in the form of a hydrodynamical shockwave \cite{Solana}. Phenomenological comparisons of this scenario with the data using the same Monte-Carlo (MC) simulation for energy loss and energy redistribution in shockwaves found agreement with both the high $P_T$ correlation pattern \cite{Dijets1,Dijets2} and the low $P_T$ peak splitting \cite{Mach1}. A comparison with the measured 3-particle correlations \cite{STAR_3p} has also been made in the same framework \cite{Mach3}, however remains somewhat inconclusive as to prove or disprove the existence of shockwaves as the chief mechanism for energy redistribution. However, in \cite{Mach2} an important difference between sonic shockwaves and other conical emission mechanisms has been pointed out, i.e. the longitudinal elongation of the shock cone due to longitudinal flow which should result in a large extension of the correlation signal in rapidity for a hydrodynamical excitation of the medium.

This elongation is obscured in single hadron triggered correlation measurements due to the fact that the rapidity of the away side parton is not determined by the rapidity of the trigger hadron and all possible rapidities of the away side parton have to be averaged. However, if the trigger is a sufficiently hard back-to-back hadron pair, then the rapidity position of the away side parton is very constrained and the elongation should be observable. Unfortunately, requiring a hard trigger hadron on the away side introduces a bias towards small energy deposition into the medium. In addition, an away side parton emerging from the medium does not only produce the leading away side hadron (which is part of the trigger) but also subleading hadrons building up correlation strength along the jet axis also at intermediate $P_T$, thus obscuring any large-angle signal of a shockwave by filling in the dip between the shockwave wings with a back-to-back peak. In this publication, we aim at a discussion of these effects.

\section{The model}

We simulate hard back-to-back hadron production in a Monte Carlo (MC) model. There are three important building blocks to this computation: 1) the primary hard parton production, 2) the propagation of the partons through the medium and 3) the hadronization of the partons. Only step 2) probes properties of the medium, and hence it is here that we must specify details of the evolution of the medium and of the parton-medium interaction. The model is described in great detail in \cite{Dijets2,dihadron-LHC}, here we will just provide a short overview.

\subsection{Primary parton production}

The production of two hard partons $k,l$ in leading order (LO) perturbative Quantum Choromdynamics (pQCD) is described by
 
\begin{equation}
\label{E-2Parton}
\frac{d\sigma^{AB\rightarrow kl +X}}{d p_T^2 dy_1 dy_2} \negthickspace = \sum_{ij} x_1 f_{i/A} 
(x_1, Q^2) x_2 f_{j/B} (x_2,Q^2) \frac{d\hat{\sigma}^{ij\rightarrow kl}}{d\hat{t}}
\end{equation}
 
where $A$ and $B$ stand for the colliding objects (protons or nuclei) and $y_{1(2)}$ is the 
rapidity of parton $k(l)$. The distribution function of a parton type $i$ in $A$ at a momentum 
fraction $x_1$ and a factorization scale $Q \sim p_T$ is $f_{i/A}(x_1, Q^2)$. The distribution 
functions are different for the free protons \cite{CTEQ1,CTEQ2} and nucleons in nuclei 
\cite{NPDF,EKS98}. The fractional momenta of the colliding partons $i$, $j$ are given by
$ x_{1,2} = \frac{p_T}{\sqrt{s}} \left(\exp[\pm y_1] + \exp[\pm y_2] \right)$.

Expressions for the pQCD subprocesses $\frac{d\hat{\sigma}^{ij\rightarrow kl}}{d\hat{t}}(\hat{s}, 
\hat{t},\hat{u})$ as a function of the parton Mandelstam variables $\hat{s}, \hat{t}$ and $\hat{u}$ 
can be found e.g. in \cite{pQCD-Xsec}. By selecting pairs of $k,l$ while summing over all allowed combinations of $i,j$, i.e. 
$gg, gq, g\overline{q}, qq, q\overline{q}, \overline{q}\overline{q}$ where $q$ stands for any of the quark flavours $u,d,s$
we find the relative strength of different combinations of outgoing partons as a function of $p_T$.

For the present investigation, we consider a dihadron trigger at midrapidity $y_1 = y_2 = 0$. By MC sampling Eq.~(\ref{E-2Parton}) we generate a back-to-back parton pair with given parton types and flavours at transverse momentum $p_T$.
To account for various effects, including higher order pQCD radiation, transverse motion of partons in the nucleon (nuclear) wave function and effectively also the fact that hadronization is not a collinear process, we fold into the distribution an intrinsic transverse momentum $k_T$ with a Gaussian distribution, thus creating a momentum imbalance between the two partons as ${\bf {p_T}_1} + {\bf {p_T}_2} = {\bf k_T}$.

\subsection{Parton propagation through the medium}

The probability density $P(x_0, y_0)$ for finding a hard vertex at the 
transverse position ${\bf r_0} = (x_0,y_0)$ and impact 
parameter ${\bf b}$ is given by the product of the nuclear profile functions as
\begin{equation}
\label{E-Profile}
P(x_0,y_0) = \frac{T_{A}({\bf r_0 + b/2}) T_A(\bf r_0 - b/2)}{T_{AA}({\bf b})},
\end{equation}
where the thickness function is given in terms of Woods-Saxon the nuclear density
$\rho_{A}({\bf r},z)$ as $T_{A}({\bf r})=\int dz \rho_{A}({\bf r},z)$. 
Rotating the coordinate system such that the near side parton propagates into the ($-x$) direction, the path of a given parton through the medium $\xi(\tau)$ is determined by its primary vertex ${\bf r_0}$ and we can compute the energy loss 
probability $P(\Delta E)_{path}$ for this path. We do this in a radiative energy loss picture \cite{Rad1,Rad2} by 
evaluating the line integrals
\begin{equation}
\label{E-omega}
\omega_c({\bf r_0}, \phi) = \int_0^\infty \negthickspace d \xi \xi \hat{q}(\xi) \quad  \text{and} \quad \langle\hat{q}L\rangle ({\bf r_0}, \phi) = \int_0^\infty \negthickspace d \xi \hat{q}(\xi)
\end{equation}
along the path where we assume the relation
\begin{equation}
\label{E-qhat}
\hat{q}(\xi) = K \cdot 2 \cdot \epsilon^{3/4}(\xi) (\cosh \rho - \sinh \rho \cos\alpha)
\end{equation}
between the local transport coefficient $\hat{q}(\xi)$ (specifying 
the quenching power of the medium), the energy density $\epsilon$ and the local flow rapidity $\rho$ with angle $\alpha$ between flow and parton trajectory \cite{Flow}. $\epsilon$ and  $\rho$ are taken from medium evolution models \cite{Hydro,Parametrized} as discussed in \cite{Dijets2}.

$\omega_c$ is the characteristic gluon frequency, setting the scale of the energy loss probability distribution, and $\langle \hat{q} L\rangle$ is a measure of the path-length weighted by the local quenching power.
We view  the parameter $K$ as a tool to account for the uncertainty in the selection of $\alpha_s$ and possible non-perturbative effects increasing the quenching power of the medium (see discussion in \cite{Dijets1}) and adjust it such that pionic $R_{AA}$ for central Au-Au collisions is described. 

Using the numerical results of \cite{QuenchingWeights}, we obtain $P(\Delta E; \omega_c, R)_{path}$ 
for $\omega_c$ and $R=2\omega_c^2/\langle\hat{q}L\rangle$ for given jet production vertex and angle $\phi$.
In the MC simulation, we first sample Eq.~(\ref{E-Profile}) to determine the vertex of origin. For a given choice of $\phi$, we then propagate both partons through the medium evaluating Eqs.~(\ref{E-omega}) and use the output to determine $P(\Delta E; \omega_c, R)_{path}$ which we sample to determine the actual energy loss of both partons in the event.

\subsection{Hadronization}

Finally, we convert the simulated partons into hadrons, provided that a back-to-back pair emerges from the medium after energy loss. More precisely, in order to determine if there is a trigger hadron above a given threshold, given a parton $k$ with momentum $p_T$, we need to sample $A_1^{k\rightarrow h}(z_1, p_T)$, i.e. the probability distribution to find a hadron $h$ from the parton $k$ where $h$ is the most energetic hadron of the shower and carries the momentum $P_T = z_1 \cdot p_T$.

In previous works \cite{Dijets1,Dijets2} we have approximated this by the normalized fragmentation function $D_{k\rightarrow h}(z, P_T)$, sampled with a lower cutoff $z_{min}$ which is adjusted to the reference d-Au data. This procedure can be justified by noting that only one hadron with $z> 0.5$ can be produced in a shower, thus above $z=0.5$ the $D_{k\rightarrow h}(z, P_T)$ and $A_1^{k\rightarrow h}(z_1, p_T)$ are  (up to the scale evolution) identical, and only in the region of low $z$ where the fragmentation function describes the production of multiple hadrons do they differ significantly.

We improve on these results by extracting $A_1(z_1, p_T)$ from the shower evolution code HERWIG \cite{HERWIG}. The procedure is described in detail in \cite{dihadron-LHC}. Sampling $A_1(z_1, p_T)$ for any parton which emerged with sufficient energy from the medium provides the energy of the two most energetic hadrons on both sides of the event. The harder of these two defines the near side. The hadron opposite to it is then the leading away side hadron. For the present investigation, we require both to be in given momentum windows to count a dihadron triggered event. We average the energy loss on near and away side parton over many such events to determine the average energy deposition into the medium. 

In order to compute the correlation strength associated with subleading fragmentation of a parton emerging from the medium we evaluate $A_2(z_1, z_2, p_T)$ (also extracted from HERWIG), the conditional probability to find the second most energetic hadron at momentum fraction $z_2$ {\em given that the most energetic hadron was found with fraction $z_1$}. This contribution to the strength of the away side correlation is competing with the shockwave signal.

Our way of modelling hadronization corresponds to an expansion of the shower development in terms of a tower of conditional probability denities $A_N(z_1, \dots, z_n, \mu)$ with the probability to produce $n$ hadrons with momentum fractions $z_1, \dots z_n$ from a parton with momentum $p_T$  being $\Pi_{i=1}^n A_i(z_1,\dots z_i,p_T)$. Taking the first two terms of this expansion is justified as long as we are interested in sufficiently hard correlations. However, in the following we  also consider situations in which the near side trigger momentum is rather hard $O(10)$ GeV, the away side trigger momentum is likewise hard $O(5)$ GeV, but with a substantial gap between near and away side to allow for energy deposition in the medium, but observe fragmentation yield associated with this trigger in a regime where hydrodynamics is valid, i.e. $O(1)$ GeV. Since the dihadron trigger forces the parton to high momenta, multi-hadron production at the low associate scale is likely. Consequently, we have to include the next terms in the expansion. A detailed numerical treatment is very complicated, however we estimate the next two terms as

\begin{equation}
A_3(z_1, z_2, z_3, p_T) \approx A_2(z_1+z_2, z_3, p_T) \theta(z_2-z_3)
\end{equation}

and

\begin{equation}
\begin{split}
A_4(z_1, z_2, z_3, z_4, p_T) \approx& A_2(z_1+z_2+z_3, z_4, p_T) \\
& \times \theta(z_2-z_3) \theta(z_3 - z_4).
\end{split}
\end{equation}

This procedure explicitly guarantees energy-momentum conservation and preserves the correct ordering in hadron momenta inside the jet. For the results quoted in the following, we have verified that the results converge and that $A_4$ is only a correction, and that hence the inclusion of further terms does not alter the result substantially.

\section{Results}

\begin{figure*}[htb]
\epsfig{file=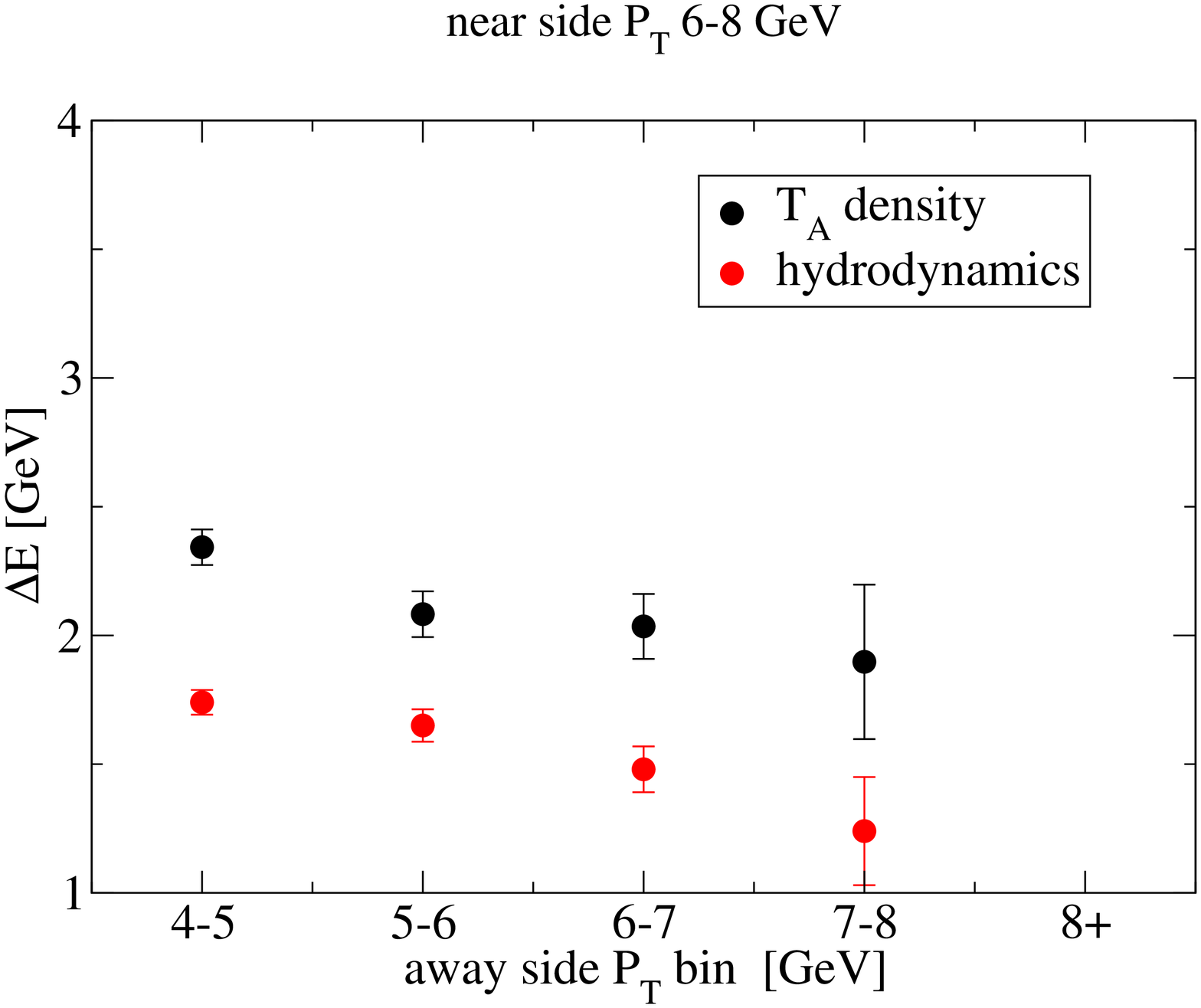, width=7.5cm}\epsfig{file=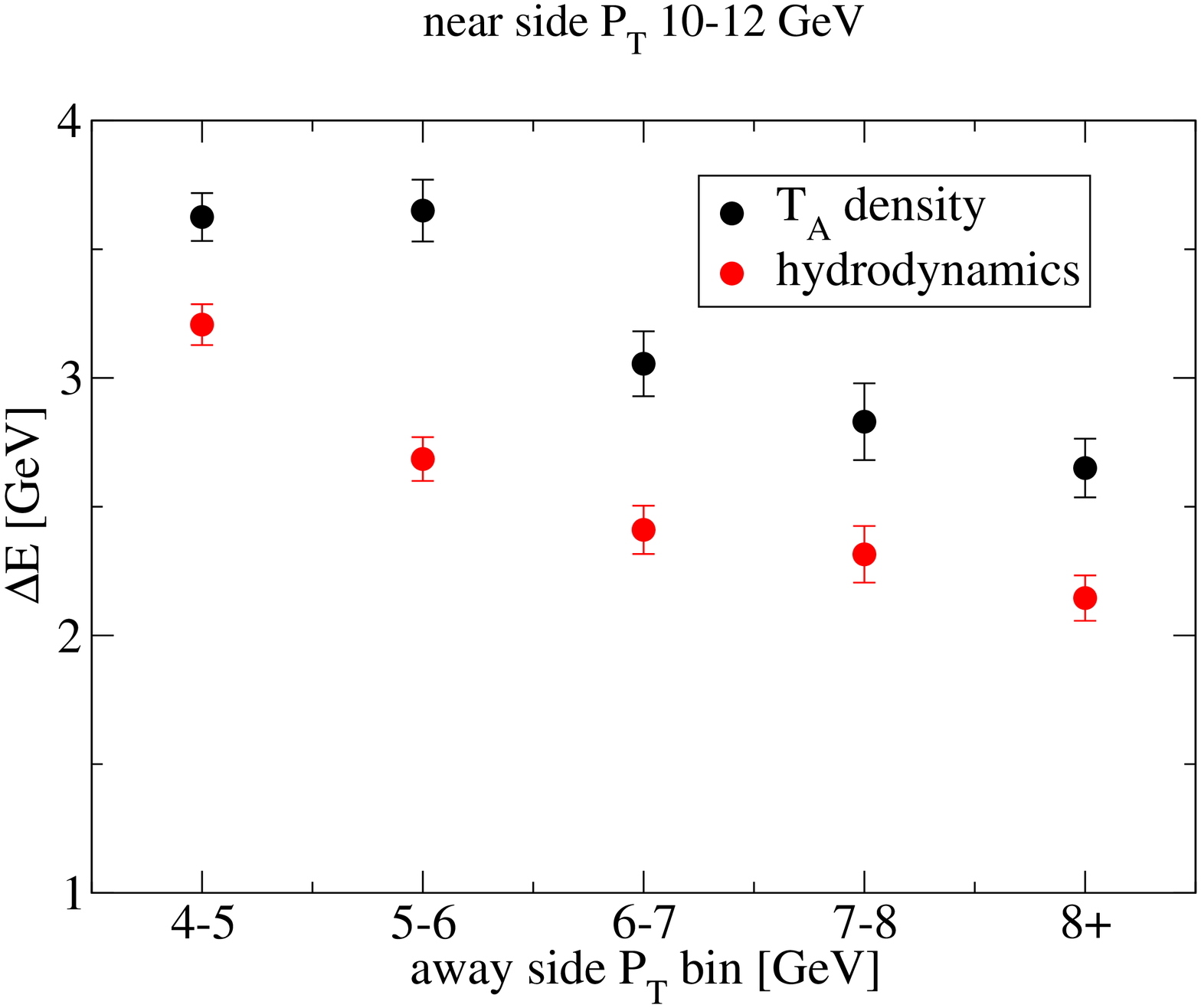, width=7.5cm}
\caption{\label{F-Edep}Average energy deposition on the away side for different dihadron trigger momentum ranges. The left panel shows 6-8 GeV momentum for the hardest hadron, the right panel 10-12 GeV. The $x-$axis shows different bins for the away side hadron momentum, the $y-$axis the corresponding energy deposition for two different models of the medium evolution (see \cite{Dijets2} for details).}
\end{figure*}

In Fig.~\ref{F-Edep} we show the away side energy deposition into the medium created in central Au-Au collisions at 200 AGeV as a function of the trigger momenta on near and away side for two different medium evolution models, a hydrdynamical code \cite{Hydro} and a parametrized evolution model \cite{Parametrized}. This is the energy available to excite a shockwave. Note that according to the phenomenological analysis \cite{Mach1,Mach3,Mach2} a large fraction $f=0.75$ (but not all) of the available energy actually excites a shockwave. 

The energy deposition is always largest when the gap between near side and away side trigger momentum is maximal. There is some dependence on what model for the medium evolution is assumed to be valid, however some general trends remain robust: The energy deposition is roughly a third of the highest (near side) trigger energy. The additional variation with the away side $P_T$ is about 50\%.

On the other hand, if {\em no} away side trigger is required, typically all of the energy of the away side parton is lost to the medium \cite{Mach1,Dijets2}. Since the parton energy is on average roughly a factor two more than the energy of the leading hadron, requiring a dihadron trigger reduces the signal strength of the shockwave by about a factor six as compared to a single hadron triggered event.

Let us now compare the strength of the shockwave correlation signal with next-to-leading and higer order fragmentation of the away side parton. For this comparison, we consider the associate momentum range of 1-2.5 GeV where the PHENIX collaboration has first seen indications for a shockwave \cite{PHENIX_Mach}. As explained in detail in \cite{Mach1,Mach3,Mach2}, we cannot reliably compute the precise magnitude of the shockwave per-trigger yield in a given momentum window, especially as long as the trigger is in a semi-hard regime below 6 GeV, as the yield is not only dependent on assumptions about flow in the medium, but also recombination/coalescence processes \cite{Coalescence,Coalescence2,Reco} need to be addressed below this scale. However, let us boldly assume that the per-trigger yield in single-hadron triggered shockwave events scales with the average trigger momentum and based on this assumption extrapolate from the PHENIX data with a trigger of 2.5-4 GeV to the two fragmentation-dominated trigger ranges of 6-8 GeV and 10-12 GeV considered in this publication (note that there is good evidence from STAR data \cite{STAR_Mach} that the rise of the yield is in fact substantially slower with trigger $P_T$). With this maximal assumption, the per trigger yield given the PHENIX acceptance in the 1-2.5 GeV associate momentum window for a 6-8 GeV trigger would be $O(2.5)$ and for a 10-12 GeV trigger $O(4)$, and, again on the level of a rough approximation, reduced down to $O(0.4)$ and $O(0.7)$ in dihadron triggered events due to the bias on energy loss.

\begin{figure}
\epsfig{file=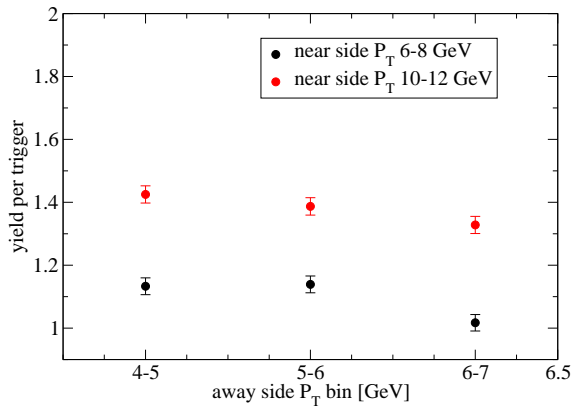, width=7.5cm}
\caption{\label{F-Frag}Estimated per-trigger yield into the 1-2.5 GeV momentum bin due subleading hadron production in fragmentation of the away side parton as a function of near side and away side $P_T$ range.} 
\end{figure}

On the other hand, the per-trigger yield into the 1-2.5 GeV associate momentum window due to subleading fragmentation of the away side parton can be computed in our hadronization scheme using the approximations for $A_3$ and $A_4$ described above. The results are summarized in Fig.~\ref{F-Frag}.

As seen from the figure, the yield is chiefly determined by the highest momentum scale (i.e. the near side trigger momentum) which is natural, given that this sets the overall energy available for hadron production in the jet. As the away side momentum scale is increased, the associated yield decreases. This is not unexpected, as requiring a larger fraction of the parton momentum to end up in the leading hadron, less momentum is available for subleading hadrons.

However, the most striking result is that the expected per-trigger yields are of order $O(1)$, i.e. they are in fact by about a factor two larger than the upper limit for the per-trigger yields caused by the medium recoil due to the shockwave. This means that if dihadron triggers are used to study shockwave production, the dominant signal at midrapidity where the away side trigger hadron is observed is not the shock cone, but rather hadrons produced in NL fragmentation processes of the trigger parton. The shockwave must then be observed as a correction to this signal. Most importantly, a splitting of the peak with a dip at zero degrees and strength at large angles is not expected under these conditions.

\section{Summary}

We have discussed the expected changes in the correlation pattern seen in a hydrodynamical momentum regime when one goes from single hadron triggered events to dihadron triggered events. The main advantage of a dihadron trigger is that the rapidity of the away side parton is tightly constrained, thus a study of the medium recoil on the away side as a function of rapidity becomes meaningful. However, there are two effects which complicate the observation of the medium recoil substantially. First, by requiring a hard away side hadron, there is a significant bias towards events in which little or no energy was deposited into the medium. This reduces the energy available to excite a shockwave, and hence the strength of the correlation by at least a factor six.

Furthermore, once a hard away side hadron is detected, it is almost unavoidable that subleading, softer hadrons are created within the shower. This contribution is rather strong at low momenta and competes with the bulk recoil of the medium. We estimated here that it is at the position of the away side parton about a factor two stronger than the medium recoil.

However, it is possible to eliminate the latter contribution due to its different shape in rapidity: While any shockwave signal is expected to be elongated in rapidity due to longitudinal flow, the jet cone due to fragmentation in vacuum would not be elongated at all. Thus, by observing associate hadron production displaced in rapdidity from a hard dihadron trigger, a (weak) shockwave signal should become visible without any contamination from soft hadron production in the jet.

\begin{acknowledgments}

I would like to thank J.~Ruppert and J.~Rak for stimulating discussions and helpful comments. This work was financially supported by the Academy of Finland, Project 115262.
 
\end{acknowledgments}

\end{document}